\documentclass[english]{article}
\usepackage[T1]{fontenc}
\usepackage[utf8]{inputenc}
\usepackage[english]{babel}
\usepackage{cite}
\usepackage{graphicx}
\usepackage[]{subfig}
\usepackage[final,markup=nocolor]{changes}

\begin{document}

\title{Surfactant films in lyotropic lamellar (and related) phases:
  Fluctuations and interactions}

\author{Frédéric Nallet \\ Université de Bordeaux. Centre de recherche
  Paul-Pascal--CNRS \\ 115 avenue du Docteur-Schweitzer \\ F-33600
  Pessac, FRANCE}

\date{\small This paper is dedicated to Dr Dominique \textsc{Langevin}
  on the occasion of her 70$^{\mathrm{th}}$ birthday}

\maketitle


\begin{abstract}
  The analogy between soap films thinning under border capillary
  suction and lamellar stacks of surfactant bilayers dehydrated by
  osmotic stress is explored, in particular in the highly dehydrated
  limit where the soap film becomes a Newton black film. The nature of
  short-range repulsive interactions between surfactant-covered
  interfaces and acting across water channels in both cases will be
  discussed.
\end{abstract}
\section{Introduction }
\label{intro}
\begin{figure}[htb]
  \centering
  \includegraphics[width=0.9\columnwidth]{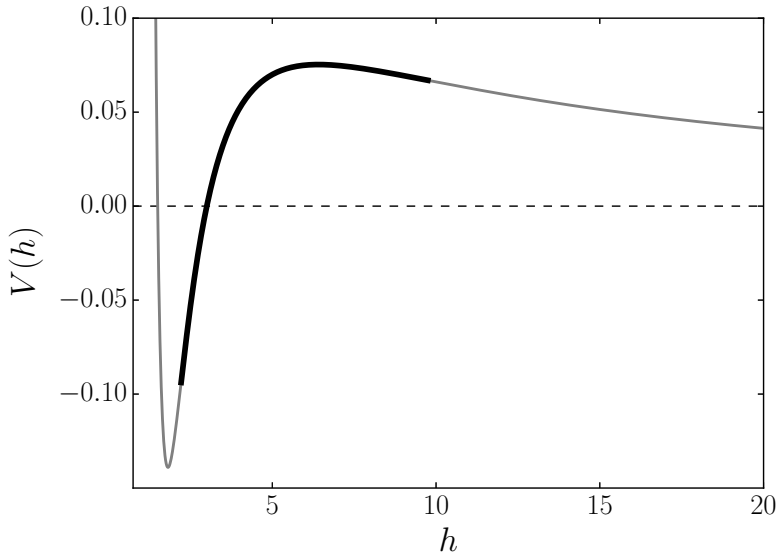}
  \caption{Interaction potential $V$ \emph{vs.} separation $h$ curve
    exhibiting a region where $\partial^2 V/\partial h^2<0$, which
    leads to a peristaltic instability. Newton (respectively common)
    black films correspond to the light-grey region of the curve on the
    left (resp. right) of the thick black region}
  \label{fig:interPot}
\end{figure}

Bubble coalescence in foams obviously involves the rupture of the
liquid film separating adjacent bubbles. However, this phenomenon
still remains not well understood in the presence of surface-active
agents~\cite{Langevin2015}. A freely-suspended soap film is a
convenient model for studying liquid films and, when drawn
horizontally, the film thins as long as capillary forces are not
opposed by the repulsive component of interactions across the water
channel. The interactions between the two surfactant-covered
air--water interfaces (with both repulsive and attractive components
in general) give rise to the so-called \emph{disjoining
  pressure}~\cite{Derjaguin1937,Derjaguin1940}. If the film does not
rupture before reaching a force cancellation point, \emph{two} states
are possible, differing by the thickness $h$ of their water
channels. They are commonly described as ``common black films'' (CBF)
or ``Newton black films'' (NBF) for the thicker or the thinner states,
respectively. Following the analysis proposed by
\textsc{Vrij}~\cite{Vrij1966}, the common black film may actually not
be a \emph{stable} state because spontaneous (thermally-excited)
thickness fluctuations around the film average thickness
$h_{\mathrm{CBF}}$--fluctuations also pictorially described as
``peristaltic fluctuations''~\cite{Young1981}--become unstable under
particular circumstances when their wavelength is larger than a
critical value $\lambda_c$.  For the instability to occur, the
repulsive component of the interaction (often considered to be of
electrostatic origin) should combine with van der Waals attractions to
yield an interaction potential $V(h)$ as a function of $h$ exhibiting
a \emph{concave} shape--see Fig.~\ref{fig:interPot}.
In other words $\partial^2 V/\partial h^2<0$ when evaluated at
$h=h_{\mathrm{CBF}}$ leads, generally speaking, to an unstable common
black film.
\par In Ref.~\cite{Vrij1966}, the critical wavelength is given by:
\begin{equation}
  \label{eq:vrij}
  \lambda_c=\left[-2\pi^2\gamma\times\frac{1}{\left.\frac{\partial^2 V}{\partial
          h^2}\right|_{h_{\mathrm{CBF}}}}\right]^{1/2}
\end{equation}
with $\gamma$ the interfacial tension. This implies that films of
finite size $L$ remain robust against peristaltic destabilisation as
long as $\lambda_c$ remains larger than $L$.
\par In cases where a common black film of thickness
$h=h_{\mathrm{CBF}}$ is unstable, and within the framework of
\textsc{Vrij}'s analysis, diverging peristaltic fluctuations lead to a
Newton black film only if $V$ changes again its concavity at some
thickness $h<h_{\mathrm{CBF}}$. This requires the presence of short
range repulsive forces between the two surfactant-covered air--water
interfaces, or a physical mechanism generating \emph{effective}
repulsive forces strong enough to overcome van der Waals
attractions. This is indeed the case for the toy model leading to the
curve $V(h)$ displayed in Fig.~\ref{fig:interPot}, but a clear
physical model for such a strong repulsion is still
lacking. Experimentally, it is observed that soap films drawn from too
\emph{dilute} surfactant solutions do \emph{not} form Newton black
films (see, for instance, Ref.~\cite{Bergeron1996}), but rupture
instead, the (qualitative) explanation being that the two facing
surfactant layers are then not dense enough for generating the
required repulsion~\cite{Bergeron1996,Langevin2015}. Conversely,
highly-loaded surfactant layers should then always lead to
\emph{stable} Newton black films, following the same reasoning. If
correct, the instability-driven CBF-to-NBF transition would then be
quite similar to \emph{spinodal decomposition}~\cite{Cahn1958}, with
common and Newton black films being analogous to the gas and liquid
phases, respectively, $\partial^2 V/\partial h^2$ being somehow
analogous to the bulk compressibility in the phase separation
analogue. The (dilute) gas phase ``collapses'' into a denser liquid
phase owing to long-range, attractive interactions but strong,
short-range repulsions prevent a further ``big crunch'' at infinite
densities. As is well known, these \deleted{such} short-range
repulsions between the molecular components of the liquid or gas
phases are actually \emph{effective} interactions, commonly explained
as arising from Pauli repulsion of overlapping electron
orbitals~\cite{Chaikin1995}. There is unfortunately no such widely
accepted explanation for the mechanism stabilising Newton black films,
authors often referring for instance to ``subtle and complex forces
associated with the structure of water''~\cite{Belorgey1991}, to
``short-range entropic confinement forces''~\cite{Bergeron1999} or to
``non-conventional interactions of entropic origin''~\cite{Jang2006}.
\par The \emph{rupture} of the Newton black film is beyond the
spinodal decomposition analogy, of course, and cannot be addressed
within the framework of \textsc{Vrij}'s analysis: As already
mentioned, there is no clear mechanism implementing another curvature
inversion in the disjoining pressure equation of state for channel
thicknesses smaller than $h_{\mathrm{NBF}}$. A key ingredient could be
taking into account \emph{surface elasticity}, because it controls the
amplitude of thermal fluctuations in the surfactant layer
density~\cite{Langevin2015,Young1981,Espert1998,Bergeron1999,Langevin2011,Rio2014}. What 
is still lacking, however, is a mechanism coupling then explicitly
surface elasticity--an ``in-plane'' property--to disjoining pressure,
acting in the direction perpendicular to the surfactant-covered
interfaces, even though at least one proposition has been formulated
in the past~\cite{Young1981}.
\par Another analogy is also quite relevant in the understanding of
soap films. The so-called lyotropic lamellar phases are prepared by
mixing appropriate amounts of lipids, soaps or surfactants with a
solvent--water, to be specific--in conditions where self-assembly
leads to the formation of bilayers of surface-active molecules,
separated by water channels and periodically stacked along an axis,
$z$, perpendicular to their average plane~\cite{Luzzati1960}. Such
system are structurally close to soap films. By controlling the
osmotic pressure $\Pi_o$ of water, as pioneered in
Ref.~\cite{LeNeveu1976}, the thickness of the water channels can be
varied and, in the limit of high pressures, strongly dehydrated
lamellar phases are obtained. Travelling along $z$ for one stacking
period $\ell$, one would go from half a bilayer to water to half a
bilayer again and thus describe the classical three-layer structure of
a soap film~\cite{Clunie1966}. When the thickness of the water core is
comparable to, or even less than the bilayer thickness, there is
little structural difference (not considering the gas phase, though)
between a Newton black film and a lamellar phase at scale
$\ell$.
\par \emph{Three-component} lamellar phases, with both oil and water
swelling each the appropriate side of surfactant monolayers, would be
even better analogues to Newton black films at simultaneously high oil
and low water contents, a situation also encountered in adhesive
oil-in-water emulsions~\cite{Poulin1996}, and rather close (upon
interchange between oil and water phases) to the so-called ``grey
films'' in water-in-oil
emulsions~\cite{Czarnecki2017}. \emph{Interactions} between
surfactant-covered oil-water interfaces, across the oil channel as
well as across the water channel, have indeed been studied (with a
different perspective, though) in such three-component lamellar
systems~\cite{Oda1997}, but \emph{not} in the limiting case of a very
thin water channel. Besides, quantitative studies of interactions in
oil-in-water emulsions when oil droplets are close to molecular
contact remain scarce~\cite{Mondain-Monval1996,Gotchev2011}.

\section{The two-component lyotropic lamellar phase}
\label{Lalpha}
From a thermodynamic point of view, the description of a two-component
system requires a single composition variable, for instance the
``solute'' (surfactant, in the present context) volume fraction
$\phi$, as it is reasonably safe to assume incompressibility. From a
structural point of view, a minimal set of two variables should also
be introduced for describing a lamellar stack of bilayers: the
stacking period $\ell$ and a surface concentration, for instance
$\Gamma$, the number of surfactant molecules per unit \emph{monolayer}
area. Within simple geometric assumptions, these three variables are
not independent. Assuming homogeneous and ideally flat bilayers, a
simple geometric description gives
\begin{equation}
  \label{eq:dil}
  \ell\phi=2v\Gamma
\end{equation}
with $v$ the volume of a surfactant molecule--a well-defined constant
quantity owing to incompressibility. If, as an additional hypothesis,
the surface concentration $\Gamma$ is constant, eq.~(\ref{eq:dil})
becomes the well-known one-dimensional ``swelling law'',
experimentally satisfied in numerous (rather dilute) systems (one
example, among many others, may be found in Ref.~\cite{Ott1990}),
sometimes up to a small logarithmic correction attributed to the
thermally-induced crumpling of the bilayers~\cite{Strey1990}.
\par It has been observed long ago, however, that $\Gamma$ experiences
significant variations in the opposite limit of dehydrated lamellar
phases. Relevant examples, among many others, may be found in
Ref.~\cite{Reiss-Husson1967,Parsegian1979,Lis1982,Lis1982a}. In
qualitative terms, this is easily understood: For a given composition
$\phi$, the ratio $\ell/\Gamma$ is constant, and a compromise should
be found between inter-bilayer interactions (acting along the stacking
axis $z$), favouring \emph{large} $\ell$ when repulsive, and
intra-bilayer (in-plane) interactions, favouring an ``optimal''
interfacial coverage $\Gamma^*$, presumably close to parameter
$\Gamma_{\infty}$ in Langmuir, Frumkin and other equations of state
describing surfactant adsorption at the air--water interface--see, for
instance, Ref.~\cite{Kralchevsky1999} for more details. Upon
dehydration, as $z$-interactions become stronger, $\ell$ ``wins'' and
the surface concentration is bound to increase above $\Gamma^*$. The
model recently proposed in Ref.~\cite{LeiteRubim2016}, recalled
briefly below, similar in spirit to the model found in
Ref.~\cite{Leontidis2007} but in a case where electrostatic
interactions are irrelevant, gives an explicit physical mechanism for
such a behaviour.
\subsection{Thermodynamic considerations}
\label{rafael}
The model starting point is \added{the core part of the Gibbs free
  energy density, \emph{viz.}} the excess \added{(Helmholtz)} free
energy density $f_{\mathrm{exc}}$ as a function of two among the three
relevant parameters $\ell$, $\phi$ and $\Gamma$, considering
eq.~(\ref{eq:dil}) to be valid. Choosing composition variables $\phi$
and $\Gamma$:
\begin{equation}
  \label{eq:fexc}
  f_{\mathrm{exc}}(\phi,\Gamma)
\end{equation}
explicit forms for the ``dilution law'', or longitudinal equation of
state $\ell(\phi)\equiv2v\Gamma_{\mathrm{opt}}(\phi)/\phi$, as well as
for the lateral equation of state $\Gamma_{\mathrm{opt}}(\phi)$ are
deduced from the minimisation equation
\begin{equation}
  \label{eq:minimisation}
  \frac{\partial f_{\mathrm{exc}}}{\partial\Gamma}=0
\end{equation}
when an explicit form for $f_{\mathrm{exc}}$ is chosen.
\par In Ref.~\cite{LeiteRubim2016}, the free energy density is built
by adding ``longitudinal'' and ``lateral'' contributions, the former
accounting for the periodic stacking, and the latter for the stacked
objects, \emph{viz.} bilayers. Owing to their two-dimensional nature,
bilayers are subjected to conformational fluctuations (or
``undulations'') with a specific character, as the mean square
amplitude of undulation fluctuations for an isolated bilayer is
divergent in the limit of infinite lateral extension. In a lamellar
phase, bilayers are stacked and large amplitude undulations are
prohibited to an extent that depends on the thickness of the water
channels. In the analysis given by
\textsc{Helfrich}~\cite{Helfrich1978}, the resulting configurational
entropy penalty is estimated as
\begin{equation}
  \label{eq:helfrich}
  f_{\mathrm{und}}=\frac{1}{\ell}\times\frac{3\pi^2}{128}\frac{\left(k_BT\right)^2}{\kappa}\frac{1}{\left(\ell-\delta\right)^2}
\end{equation}
(here written with a $-T$ factor to become a free energy density) in
terms of bilayer bending modulus $\kappa$ and bilayer thickness
$\delta$. Note that, similarly to the \textsc{van der Waals}
``correction'' to the ideal gas free energy accounting for excluded
volume effects, with a divergence in the high density limit,
eq.~(\ref{eq:helfrich}) diverges in the wholly dehydrated limit where
$\ell\rightarrow\delta$ and that, when dealing with non-ideal gaseous
phases, this feature has been criticised long
ago~\cite{Clerk-Maxwell1874}.
\par In lamellar phases, interactions between bilayers, including van
der Waals attractive forces of course, have to be taken into account,
as already recognised in Ref.~\cite{Helfrich1978}. But, as argued by
\textsc{Milner} and \textsc{Roux}~\cite{Milner1992}, one should
\emph{not} merely add to the entropic term in the free energy the
interaction potentials relevant for the forces being considered, but
rather incorporate their effects as a \emph{virial} term. This amounts
to adding a term
\begin{equation}
  \label{eq:viriel}
  f_{\mathrm{vir}}=-k_BT\chi\phi^2
\end{equation}
where $\chi$ measures the average strength of the inter-bilayer
interactions across the whole range of possible separations, similarly
to the measure provided by the second virial coefficient for
inter-molecular interactions in the classical theory of liquid--gas
phase separation.
\par As far as the ``lateral'' contribution to the free energy is
concerned, inspiring ideas may be found by considering the related
problem of the equation of state of surface-adsorbing molecules (or
particles) accounting for both excluded area and in-plane
interactions, as discussed--for instance--in
Ref.~\cite{Chatelier1996,Kralchevsky1999} and \cite{Groot2010},
respectively. For the present purpose, the ``van der Waals'' approach
has been chosen, with the corresponding free energy expression
\begin{equation}
  \label{eq:vdw2D}
  f_{\mathrm{2D}}=\frac{k_BT}{\ell}\times\left\{\Gamma\ln\left(\frac{\Gamma}{\Gamma_{\infty}-\Gamma}\right)+\frac{1}{2}b_2\Gamma^2\right\}
\end{equation}
parameters $\Gamma_{\infty}$ and $b_2$ describing, respectively, the
upper limit for surfactant adsorption (excluded area) and, in terms of
a second virial coefficient, in-plane (cohesive) interactions.
\par With the above-described physical ingredients, the model for the free
energy density becomes~\cite{LeiteRubim2016}:
\begin{eqnarray}
  \label{eq:rafael}
  f_{\mathrm{exc}}&=&\frac{1}{\ell}\times\frac{3\pi^2}{128}\frac{\left(k_BT\right)^2}{\kappa}\frac{1}{\left(\ell-\delta\right)^2}
  \nonumber\\ 
  &&-k_BT\chi\phi^2+\frac{k_BT}{\ell}\times\left\{\Gamma\ln\left(\frac{\Gamma}{\Gamma_{\infty}-\Gamma}\right)+\frac{1}{2}b_2\Gamma^2\right\}
\end{eqnarray}
which is also, introducing reduced quantities:
\begin{eqnarray}
  \label{eq:rafaelReduced}
\tilde{f}_{\mathrm{exc}}&=&\frac{3\pi^2}{128}\times\frac{1}{\tilde{\kappa}}\times\frac{\phi^3}{\tilde{\Gamma}^3\left(1-\phi\Gamma_0/\tilde{\Gamma}\right)^2}
  \nonumber\\ 
  &&-\tilde{\chi}\phi^2+\frac{\phi}{2}\left\{\ln\left(\frac{\tilde{\Gamma}}{1-\tilde{\Gamma}}\right)+\frac{1}{2}b_2\tilde{\Gamma}\right\}  
\end{eqnarray}
with $\tilde{f}_{\mathrm{exc}}\equiv vf_{\mathrm{exc}}/(k_BT)$,
$\tilde{\Gamma}\equiv\Gamma/\Gamma_{\infty}$, 
$\tilde{\kappa}\equiv8v^2\Gamma_{\infty}^3\kappa/(k_BT)$,
$\Gamma_0\equiv\delta/(2v\Gamma_{\infty})$, $\tilde{\chi}\equiv\chi v$
and $\tilde{b}_2\equiv b_2\Gamma_{\infty}$.
\par Implicit in the four-parameter model given above, possible
dependences of $\tilde{\kappa}$ on $\tilde{\Gamma}$, and of
$\tilde{\chi}$ on $\phi$ \added{may be introduced}. On the other hand,
even if nothing prevents in \textsc{Helfrich}'s analysis
$\Gamma_0\neq1$, it is presumably legitimate to assume $\Gamma_0=1$,
thereby reducing the number of model parameters to three. The most
simple case, \emph{viz.}  $\tilde{\kappa}$ and $\tilde{\chi}$
constant, as well as $\Gamma_0=1$, will now be discussed.
\subsubsection{In-plane elasticity}
\label{surfElasticity}
\begin{figure}[htb]
  \centering
  \includegraphics[width=0.9\columnwidth]{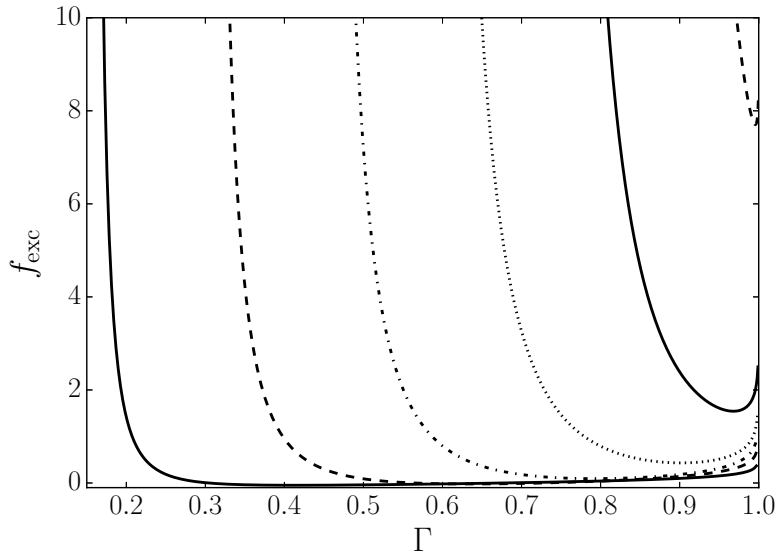}  
  \caption{Model free energy density as a function of surface
    coverage--dilution effect. Surfactant volume fractions $\phi$:
    0.15 (solid line), 0.29 (dashed line), 0.43 (dash-dotted line),
    0.57 (dotted line), 0.71 (solid line) and 0.85 (dashed
    line). Model parameters: $\kappa=1.$, $\chi=1.55$, $b_2=-1.$}
  \label{fig:surfaceCoveragePhi}
\end{figure}

A first direct consequence of the above-described model concerns the
optimal surface coverage, solving eq.~(\ref{eq:minimisation}) for
$\Gamma_{\mathrm{opt}}$. Fig.~\ref{fig:surfaceCoveragePhi} displays
$f_{\mathrm{exc}}$ as a function of $\Gamma$,
eq.~(\ref{eq:rafaelReduced})--note that \emph{reduced} units are from
now on implicitly assumed--, for representative values of the bilayer
volume fractions and arbitrarily chosen, but \emph{fixed} values of
the model parameters $\kappa, \chi$ and $b_2$.
The displacement towards higher optimal surface coverage upon
dehydrating the lamellar stack evidences the ``$\ell$--$\Gamma$''
interplay, eq.~(\ref{eq:dil}), with stronger interactions along the
stacking axis $z$ more easily overcoming in-plane interactions.
\par
\begin{figure}[htb]
  \centering
  \includegraphics[width=0.9\columnwidth]{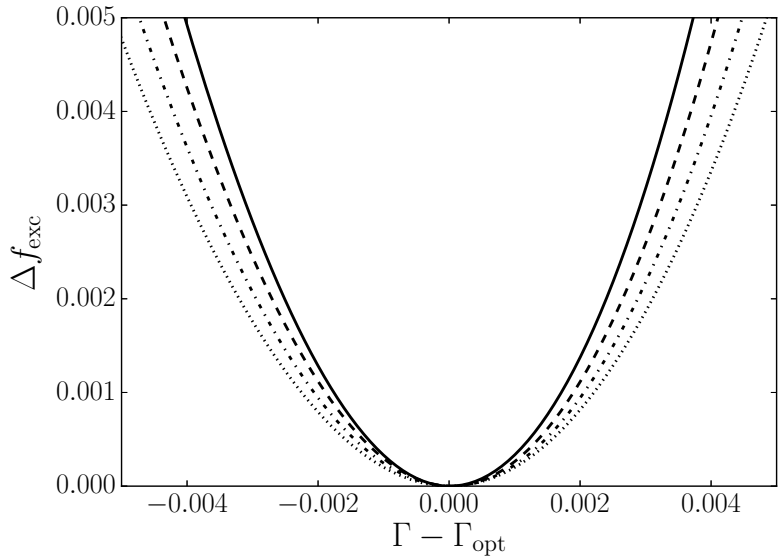}
  \caption{Model free energy density as a function of surface coverage
    relative to its optimal value $\Gamma_{\mathrm{opt}}$--in-plane
    interaction effect. Bilayer volume fraction $\phi=0.70$. Second
    virial coefficient $b_2$, from higher to lower bilayer in-plane
    cohesion: $-31$ (solid line), $-21$ (dashed line), $-11$
    (dash-dotted line) and $-1$ (dotted line). Bending modulus
    $\kappa=1.$, \textsc{Milner-Roux} virial coefficient
    $\chi=1.55$. $\Gamma_{\mathrm{opt}}\approx0.97$. Free energy
    densities translated vertically to ensure a value $0$ for
    $\Gamma=\Gamma_{\mathrm{opt}}$}
  \label{fig:surfaceCoverageB2}
\end{figure}

 As apparent in Fig.~\ref{fig:surfaceCoveragePhi}, dilution not
only changes the optimal surface coverage $\Gamma_{\mathrm{opt}}$, but
also surface \emph{elasticity} as an indirect consequence of the
``$\ell$--$\Gamma$'' interplay. The $f_{\mathrm{exc}}(\phi,\Gamma)$
curves are indeed much flatter in the vicinity of
$\Gamma_{\mathrm{opt}}$ at higher hydration. As made clear in
Fig.~\ref{fig:surfaceCoverageB2}, however, the most \emph{direct}
role--as far as surface elasticity is concerned--should be attributed
to \emph{in-plane} interactions, monitored by the cohesion parameter
$b_2$.
While the optimal surface coverage practically does not depend on
bilayer cohesion ($\Gamma_{\mathrm{opt}}$ varies by less than 0.3\%
for the conditions chosen in Fig.~\ref{fig:surfaceCoverageB2}), the
free energy density \emph{curvature} is significantly altered.
\par While the inter-bilayer interaction parameter $\chi$ has no
effect on surface elasticity (from eq.~(\ref{eq:rafaelReduced}),
$\partial^2f_{\mathrm{exc}}/\partial\Gamma^2$ does not depend on
$\chi$, obviously), the bilayer bending modulus $\kappa$ plays some
role in this respect, though less directly than $b_2$, and again
through the ``$\ell$--$\Gamma$'' interplay: A \emph{smaller} value for
$\kappa$ indeed implies a \emph{larger} configurational entropy
penalty, resulting in a stronger \emph{effective} inter-bilayer
repulsion. This leads in turn to a denser interfacial coverage and,
eventually, to a higher surface elasticity.
\subsubsection{Water activity in lamellar stacks of bilayers}
\label{muW}
The significance of parameter $\chi$ is most strikingly evidenced when
considering the so-called ``swelling limit'' of the lamellar stack. As
long as (direct or effective) inter-bilayer interactions are repulsive
enough, adding water to the system will result in swelling the
structure: $\ell(\phi)$ increases with decreasing $\phi$. This was
indeed the case for the parameters chosen in drawing
Fig.~\ref{fig:surfaceCoveragePhi}, again referring to the
``$\ell$--$\Gamma$'' interplay. Conversely, for too weak inter-bilayer
repulsions, or in the presence of strong enough \emph{attractions}, a
swelling limit may be reached: Any amount of water added to the
lamellar stack above this limit does not incorporate within the
structure but phase-separates instead as excess water. A quite
interesting property from a theoretical point of view is the presence
of a so-called ``unbinding''
transition~\cite{Lipowski1986,Podgornik1992,Milner1992}, occurring at
a \emph{finite} $\chi_u$: For \emph{unbound} lamellar systems,
\emph{i.e.} $\chi<\chi_u$, $\ell\rightarrow+\infty$ upon swelling, but
$\ell$ cannot exceed a limiting value $\ell_{\mathrm{MAX}}$ (and
excess water is expelled from the ``bound'' lamellar structure at
hydration $\phi_{\mathrm{min}}$) if $\chi>\chi_u$.
\par
\begin{figure}[htb]
  \centering
  \includegraphics[width=0.9\columnwidth]{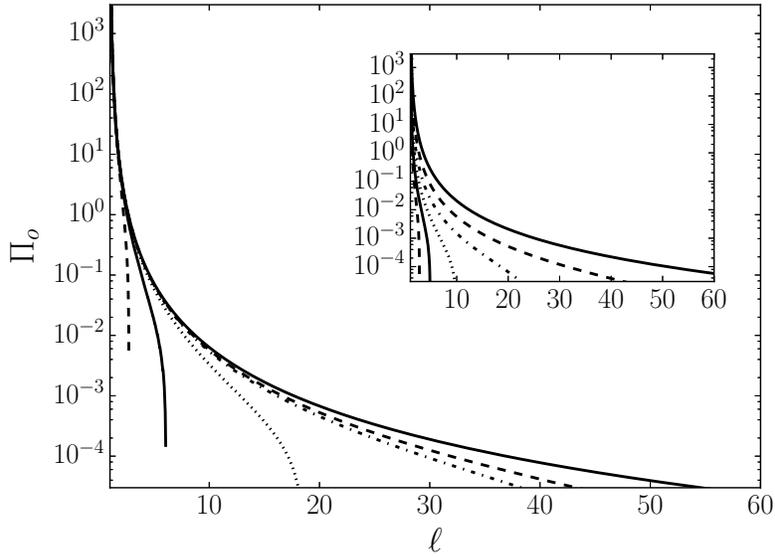}
  \caption{Osmotic pressure equation of state expressed as a function
    of the stacking period $\ell$.  \textsc{Milner}--\textsc{Roux}
    virial parameter $\chi$, from weaker to stronger inter-bilayer
    attractive interactions: 0 (upper solid line), 0.1 (upper dashed
    line), 0.15 (dash-dotted line), 0.5 (dotted line), 2. (lower solid
    line) and 8. (lower dashed line). With the values chosen for
    parameters $\kappa$ and $b_2$ (respectively, 0.1 and -11.),
    $\chi=0.15$ is close to, but greater than $\chi_u$ where unbinding
    occurs, and the swelling limit is
    $\phi_{\mathrm{min}}\approx1.4$\%. Inset: $\chi=0.11$, $b_2=-11.$,
    bilayer bending modulus $\kappa$: 0.03 (upper solid line), 0.1
    (upper dashed line), 0.3 (dash-dotted line, ``bound'' system with
    swelling limit $\ell_{\mathrm{MAX}}\approx26.6$), 1. (dotted
    line), 3. (lower solid line) and 10.  (lower dashed line)}
  \label{fig:pOsm}
\end{figure}

 For analysing the occurrence of a ``bound'' system, \emph{osmotic
  pressure} $\Pi_o$ is a relevant tool since $\Pi_o=0$ when the
swelling limit is reached. From the general thermodynamic relation
\begin{equation}
  \label{eq:pOsmDef}
  \Pi_o=\phi\frac{\partial f_{\mathrm{exc}}}{\partial\phi}-f_{\mathrm{exc}}
\end{equation}
osmotic pressure is straightforwardly expressed from
eq.~(\ref{eq:rafaelReduced}) as
\begin{equation}
  \label{eq:pOsm}
  \Pi_o=\frac{3\pi^2}{64\kappa}\frac{\phi^3}{(\Gamma-\phi)^3}-\chi\phi^2
\end{equation}
Fig.~\ref{fig:pOsm} illustrates the control that $\chi$ exerts on the
dilution limit.
Taking into account the overall trend favouring swelling that
originates in the \textsc{Helfrich} term--the first term in the
right-hand side of eq.~(\ref{eq:pOsm})--, weaker \emph{attractive}
interactions between stacked bilayers, \emph{viz.} smaller (positive)
values for the \textsc{Milner}--\textsc{Roux} virial parameter $\chi$
lead to systems able to incorporate \emph{more} water (and even an
infinite amount in the case of ``unbound'' systems).
\par The unbinding transition may of course also be reached by tuning
\emph{effective} repulsions, that is to say varying $\kappa$, instead
of playing with direct interactions through parameter $\chi$, see
inset in Fig.~\ref{fig:pOsm}.
\par An other feature of Fig.~\ref{fig:pOsm} is worth mentioning here:
In the strongly \emph{dehydrated} limit where $\phi$ is close to 1,
with therefore $\Gamma_{\mathrm{opt}}\rightarrow1$ (as suggested in
Fig.~\ref{fig:surfaceCoveragePhi}) and $\ell\rightarrow1$, the
dependence of osmotic pressure on $\ell$ is close to being
exponential. The model, eq.~(\ref{eq:rafaelReduced}), is thus able to
reproduce a property first experimentally established in pioneering
reports, Ref.~\cite{LeNeveu1976,Parsegian1979}, and often discussed in
terms of ``hydration interactions'' between surfactant bilayers (see,
for instance, Ref.~\cite{Lis1982,Leontidis2007,Parsegian2011}). In the
present model, however, no such \emph{ad hoc} interactions have been
taken into consideration: Eq.~(\ref{eq:helfrich}), (\ref{eq:viriel}),
as well as (\ref{eq:vdw2D}) incorporate only fundamental thermodynamic
ingredients \emph{without} postulating any specific profile,
\emph{e.g.} exponential, for inter-bilayer interactions. The physical
reality of the so-called ``hydration interactions'' may therefore be
considered as highly dubious~\cite{LeiteRubim2016}.
\subsection{Phase diagrams}
\label{diaPha}
As recalled in Section~\ref{intro} the CBF-to-NBF transition is a kind
of spinodal decomposition, the instability being driven within the
framework of \textsc{Vrij}'s analysis by diverging ``peristaltic''
fluctuations. An important question is whether a similar instability
actually occurs in the model, eq.~(\ref{eq:rafaelReduced}). The
competition in terms of the swelling properties between $\kappa$ and
$\chi$ would then not only lead in some cases to the ``unbinding''
transition discussed in Section~\ref{muW} above, that is to say to a
lamellar phase -- solvent phase separation but also, in other regions
of the parameter space, to a \emph{lamellar}--\emph{lamellar} phase
equilibrium. Such phase separations, more frequent in ternary or
higher multi-component systems of stacked bilayers, are indeed
sometimes observed experimentally in a few \emph{binary} lyotropic
lamellar
phases~\cite{Dubois1991,Zemb1993,McGrath1996,Silva2007,Silva2010}.
\par The so-called ``peristaltic'' fluctuations in the thin liquid
film context, associated to fluctuations in the thickness of the water
channel separating the two facing, surfactant-covered air--water
interfaces, should not be confused with homonym fluctuations in the
context of lyotropic lamellar phases, however. Presumably because the
focus is on \emph{bilayer} objects in the latter case, the ``membrane
peristaltic'' modes, according to the terminology found in
Ref.~\cite{Nallet1989}, refer to fluctuations in \emph{surface
  coverage} $\Gamma$, the fluctuations in water channel thickness
giving rise to distinct, ``baroclinic'' modes. Notwithstanding a
perhaps confusing terminology, and similarly to (thin films)
``peristaltic'' fluctuations, ``baroclinic'' fluctuations may be
\emph{divergent} in lyotropic lamellar phases. When this is the case,
a lamellar--lamellar phase separation occurs as a consequence of the
associated spinodal instability.
\par The criterion for spinodal instability is, at least formally,
rather easily established. For a lamellar stack with an overall
bilayer content given by $\phi$, and therefore a surface coverage
$\Gamma_{\mathrm{opt}}$ deduced from eq.~(\ref{eq:minimisation}),
\emph{local} fluctuations $\phi+\delta\phi(\mathbf{x})$ and
$\Gamma_{\mathrm{opt}}+\delta\Gamma(\mathbf{x})$ in composition and
surface coverage, respectively, are randomly generated by thermal
motion within a (macroscopically) infinitesimal volume $\delta\!V$
located at position $\mathbf{x}$ within the system. Using Boltzmann
statistics, such fluctuations occur with a probability $p$
proportional to
\begin{equation}
\label{eq:boltzmann}
\exp\left\{-\frac{\delta\!Vf_{\mathrm{exc}}\left[\phi+\delta\phi(\mathbf{x}),\Gamma_{\mathrm{opt}}+\delta\Gamma(\mathbf{x})\right]-\mu_w\left[N_w+\delta\!N_w(\mathbf{x})\right]-\mu\left[N+\delta\!N(\mathbf{x})\right]}{k_BT}\right\}
\end{equation}
since the (macroscopic) volume outside $\delta\!V$ plays the role of
a reservoir with chemical potentials $\mu$ and $\mu_w$ for surfactant
and solvent, respectively. Owing to incompressibility, composition
fluctuations inside $\delta\!V$ correspond to variations in the
numbers of solvent ($N_w$) and surfactant ($N$) molecules with
\emph{opposite} sign:
$v_w\delta\!N_w(\mathbf{x})+v\delta\!N(\mathbf{x})=0$, $v_w$ being the
solvent molecular volume, which implies
$\delta\phi(\mathbf{x})\propto\delta\!N(\mathbf{x})$. For ensuring
that the overall composition is indeed given by volume fraction
$\phi$, the chemical potentials should satisfy
\begin{equation}
  \label{eq:minimisation2}
  \frac{\mu}{v}-\frac{\mu_w}{v_w}=\partial
  f_{\mathrm{exc}}/\partial\phi
\end{equation}
and eq.~(\ref{eq:minimisation}) is of course recovered for ensuring
that the surface coverage is $\Gamma_{\mathrm{opt}}$.
\par For \emph{small} enough fluctuations, the probability $p$ reduces
to
\begin{equation}
\label{eq:gaussian}
p\propto\exp\left\{-\frac{\frac{\partial^2f_{\mathrm{exc}}}{\partial\phi^2}\delta\phi^2+2\frac{\partial^2f_{\mathrm{exc}}}{\partial\phi\partial\Gamma}\delta\phi\delta\Gamma+\frac{\partial^2f_{\mathrm{exc}}}{\partial\Gamma^2}\delta\Gamma^2}{2k_BT}\delta\!V\right\}
\end{equation}
truncating the Taylor expansion to second order. Stability of the
homogeneous state requires
$p\left[\delta\phi(\mathbf{x}),\delta\Gamma(\mathbf{x})\right]$ to be
smaller than
$p\left[\delta\phi(\mathbf{x})=0,\delta\Gamma(\mathbf{x})=0\right]$
for any fluctuations
$\left[\delta\phi(\mathbf{x}),\delta\Gamma(\mathbf{x})\right]$ or,
equivalently
\begin{equation}
  \label{eq:stability}
  \left\{
    \begin{array}{ccl}
      0&<&\frac{\partial^2f_{\mathrm{exc}}}{\partial\phi^2} \\
      0&<&\frac{\partial^2f_{\mathrm{exc}}}{\partial\phi^2}\frac{\partial^2f_{\mathrm{exc}}}{\partial\Gamma^2}-\left(\frac{\partial^2f_{\mathrm{exc}}}{\partial\phi\partial\Gamma}\right)^2
\end{array}
    \right.
\end{equation}
Conversely, a spinodal decomposition and, ultimately, a
lamellar--lamellar phase separation occur if one (or both)
inequalities are violated.
\par
\begin{figure}[htb]
  \centering
  \subfloat[][$\phi$ -- $\chi$ projection]{\includegraphics[width=0.45\columnwidth]{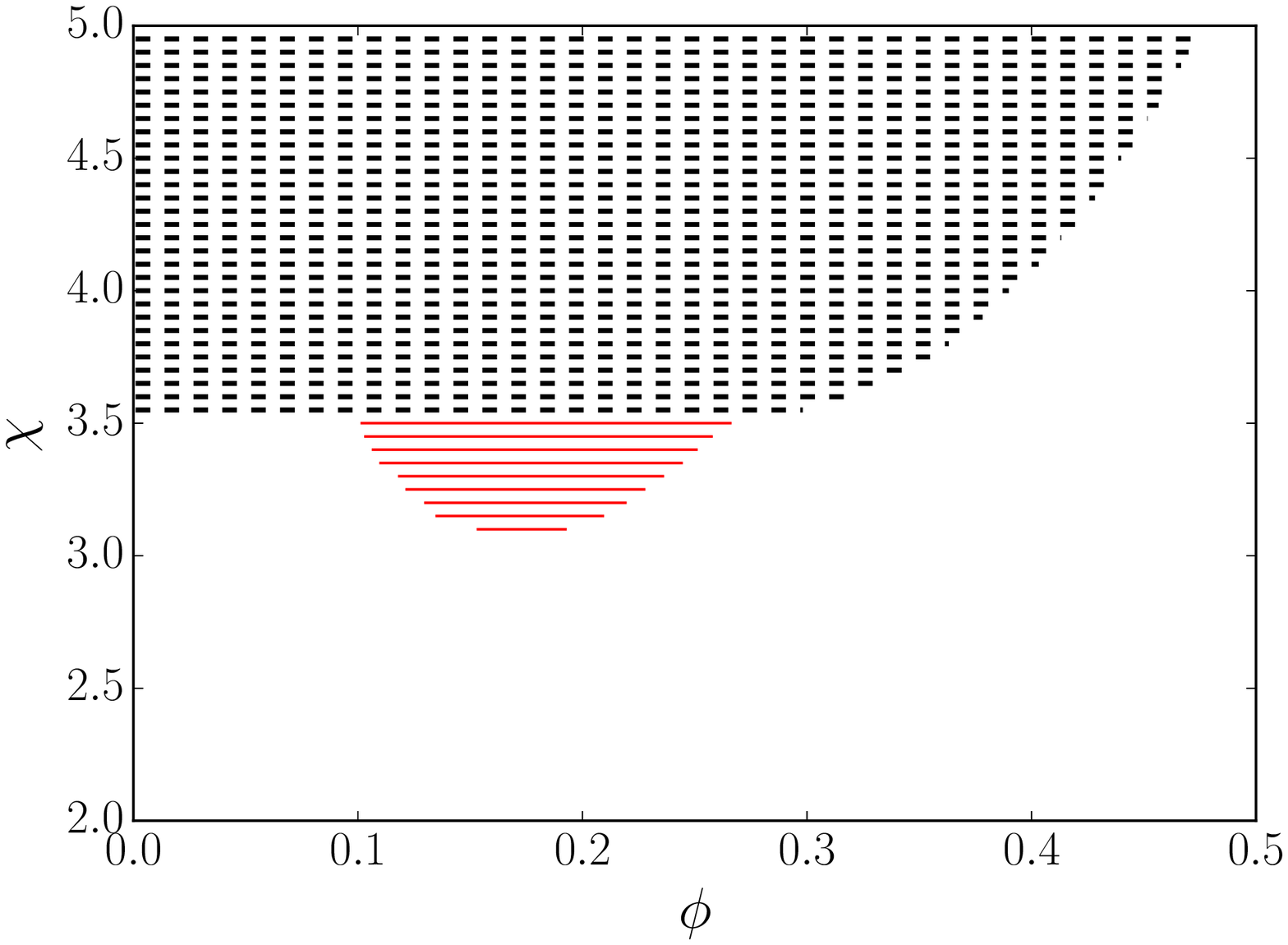}\label{fig:diaPhaChiPhi}}
\qquad\subfloat[][$\phi$ -- $\kappa$ projection]{\includegraphics[width=0.45\columnwidth]{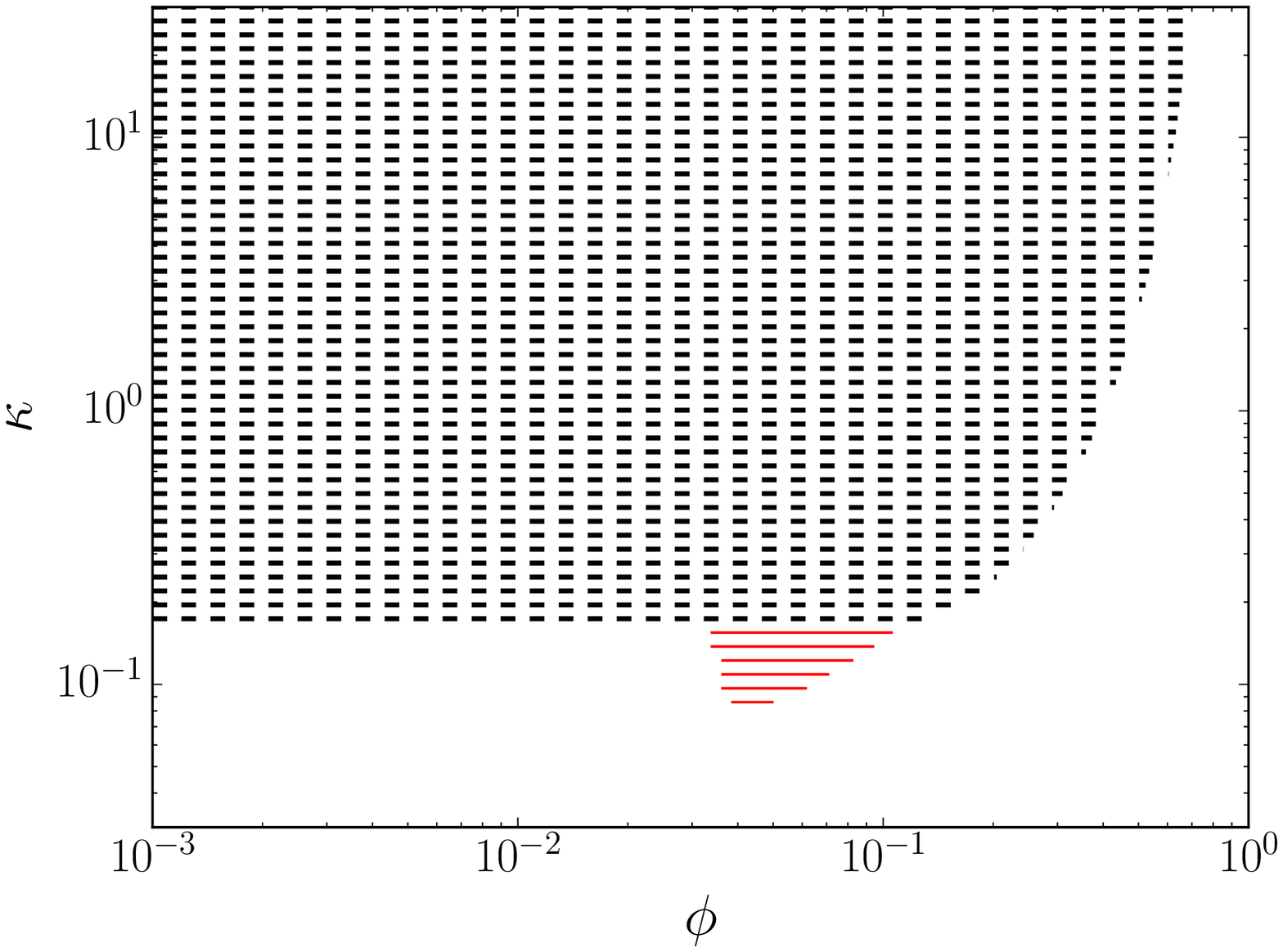}\label{fig:diaPhaKappaPhi}}
\caption{Schematic phase diagrams resulting from the model,
  eq.~(\ref{eq:rafaelReduced}). The black, dashed regions are obtained
  in analysing the coexistence between pure solvent, $\phi=0$, and a
  \emph{bound} lamellar phase with composition $\phi_{\mathrm{min}}$,
  where $\Pi_o(\phi_{\mathrm{min}})=0$. The red lines mark spinodal
  instability regions where
  $\frac{\partial^2f_{\mathrm{exc}}}{\partial\phi^2}\frac{\partial^2f_{\mathrm{exc}}}{\partial\Gamma^2}-
  \left(\frac{\partial^2f_{\mathrm{exc}}}{\partial\phi\partial\Gamma}\right)^2<0$. The
  phase boundaries for lamellar--lamellar phase separations enclose
  the red-marked regions, that end at \emph{critical points}. (a)
  Projection in the $\phi$--$\chi$ plane, with $\kappa=1.$ and
  $b_2=-0.25$. For $\chi$ smaller than $\chi_u\approx3.53$, the
  lamellar phase is unbound.  The swelling limit $\phi_{\mathrm{min}}$
  of bound systems increases with \textsc{Milner-Roux} virial
  coefficient $\chi$. Critical point coordinates $\chi_c\approx3.067$,
  $\phi_c\approx0.162$. (b) Projection in the $\phi$--$\kappa$ plane,
  with $\chi=3.$ and $b_2=-5.0$. For $\kappa$ smaller than
  $\kappa_u\approx0.2$, the lamellar phase is unbound. The swelling
  limit $\phi_{\mathrm{min}}$ of bound systems increases with bilayer
  bending modulus $\kappa$. Critical point coordinates
  $\kappa_c\approx0.0802$, $\phi_c\approx0.0415$}
  \label{fig:diaPha}
\end{figure}

 A somehow similar discussion has been outlined before, in
particular in the more complicated case of \emph{three-component}
(polymer-doped) lamellar
phases~\cite{Ligoure1997,Porcar1997,Ficheux2001}. Much more recently,
spinodal instabilities in two-component lyotropic lamellar systems
have been investigated quite in depth,
however~\cite{Bougis2016}. Considering simultaneously the spinodal
instability criteria, eq.~(\ref{eq:stability}), and the condition for
unbound swelling $\Pi_o>0$, core features of the phase diagram of
binary lamellar stacks described by eq.~(\ref{eq:rafaelReduced}) can
be established, as in the examples displayed in
Fig.~\ref{fig:diaPhaChiPhi} and \ref{fig:diaPhaKappaPhi}, illustrating
$\chi$ and $\kappa$ effects, respectively.
\par In order to actually describe the lamellar--lamellar phase
separation that is bound to occur as the outcome of the spinodal
decomposition and, therefore, to build less schematic phase diagrams
than displayed in Fig.~\ref{fig:diaPha}, the free energy associated to
two coexisting lamellar phases, \emph{i.e.}
\begin{equation}
\label{eq:two-phase}
F_{2}\propto\varphi f_{\mathrm{exc}}\left(\phi_1,\Gamma_1\right)+(1-\varphi)f_{\mathrm{exc}}\left(\frac{\phi-\varphi\phi_1}{1-\varphi},\Gamma_2\right)
\end{equation}
where $\left\{\phi_1,\Gamma_1\right\}$ and
$\left\{\phi_2\equiv(\phi-\varphi\phi_1)/(1-\varphi),\Gamma_2\right\}$
are the respective compositions variables for phases~1 and 2, $\phi$
is the overall surfactant volume fraction, and $\varphi$ the fraction
of the total volume occupied by phase~1, has to be minimised with
respect to all the fluctuating variables, namely $\Gamma_1$,
$\Gamma_2$, $\phi_1$ and $\varphi$. The procedure formally leads to
two equations similar to eq.~(\ref{eq:minimisation}), as well as two
equations stipulating osmotic equilibrium between the two phases,
namely equal surfactant chemical potentials and equal osmotic
pressures (or solvent chemical potentials). With the lever
rule $\varphi\phi_1+(1-\varphi)\phi_2=\phi$ implicit, the two latter
equations are commonly cast in the following form
\begin{equation}
  \label{eq:commonT}
  \begin{array}{lcl}
    \left.\frac{\partial
        f_{\mathrm{exc}}}{\partial\phi}\right|_{\left\{\phi_1,\Gamma_1\right\}}&=&\left.\frac{\partial
        f_{\mathrm{exc}}}{\partial\phi}\right|_{\left\{\phi_2,\Gamma_2\right\}}
    \\
    f_{\mathrm{exc}}(\phi_1,\Gamma_1)&=&f_{\mathrm{exc}}(\phi_2,\Gamma_2)+(\phi_1-\phi_2)\left.\frac{\partial
        f_{\mathrm{exc}}}{\partial\phi}\right|_{\left\{\phi_2,\Gamma_2\right\}}
\end{array}
\end{equation}
that would be geometrically interpreted as the ``common tangent
construct'' in the classical problem of a liquid--liquid phase
separation in binary solutions.
\par
\begin{figure}[htb]
  \centering
  \includegraphics[width=0.95\columnwidth]{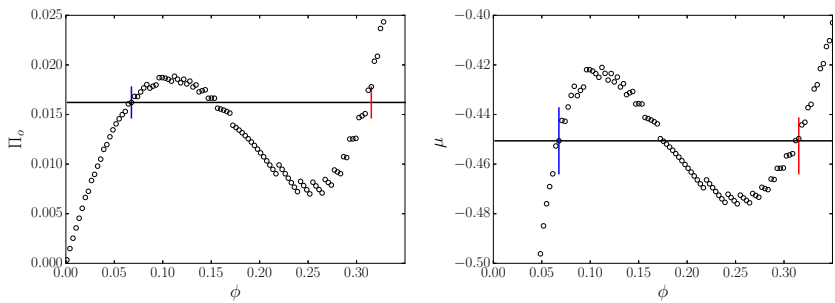}
  \caption{Osmotic pressure $\Pi_o$ and surfactant chemical potential
    $\mu$ \emph{vs.} composition $\phi$ according to numerical
    estimates from the model,
    eq.~(\ref{eq:rafaelReduced}). Parameters: $\kappa=1.$,
    $b_2=-0.25$, $\chi=3.4$. A phase separation occurs between a
    \emph{dilute} lamellar phase with composition
    $\phi_{\mathrm{dil}}\approx6.8$\% (vertical blue line. Stacking
    period $\ell_{\mathrm{dil}}\approx3.4$) and a \emph{concentrated} one with
    $\phi_{\mathrm{conc}}\approx31.2$\% (vertical red line. Stacking
    period $\ell_{\mathrm{conc}}\approx2.1$)}
  \label{fig:piMu}
\end{figure}

 In practice, considerable efforts have to be spent for explicitly
minimising eq.~(\ref{eq:two-phase}) and eventually get phase
boundaries $\phi_1$ and $\phi_2$, as well as surface coverages
$\Gamma_{\mathrm{opt}}(\phi_1)$ and $\Gamma_{\mathrm{opt}}(\phi_2)$.
The problem is indeed more difficult than \emph{separately} solving
for the volume fraction $\phi$ the equations $\Pi_o=0$,
$\partial^2f_{\mathrm{exc}}/\partial\phi^2=0$ and
$\partial^2f_{\mathrm{exc}}/\partial\phi^2\times\partial^2f_{\mathrm{exc}}/\partial\Gamma^2-\left(\partial^2f_{\mathrm{exc}}/\partial\phi\partial\Gamma\right)^2=0$.
Fig.~\ref{fig:piMu} gives an illustration of the \emph{simultaneous}
solution for \emph{two} volume fractions of eq.~(\ref{eq:commonT}) in
terms of osmotic pressure and surfactant chemical potential.
\par
\begin{figure}[htb]
  \centering
  \includegraphics[width=0.95\columnwidth]{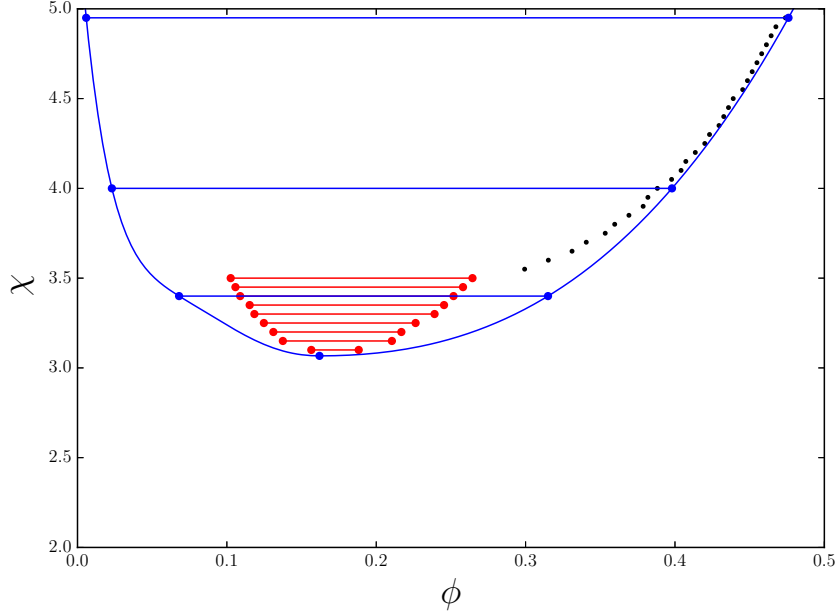}
  \caption{Projection in the $\phi$--$\chi$ plane, with $\kappa=1.$
    and $b_2=-0.25$ of the phase diagram resulting from the model,
    eq.~(\ref{eq:rafaelReduced}). The black circles mark loci where
    the osmotic pressure $\Pi_o(\phi)$ first vanishes when increasing
    hydration. The red circles mark loci where
    $\Delta\equiv\frac{\partial^2f_{\mathrm{exc}}}{\partial\phi^2}\frac{\partial^2f_{\mathrm{exc}}}{\partial\Gamma^2}-
    \left(\frac{\partial^2f_{\mathrm{exc}}}{\partial\phi\partial\Gamma}\right)^2$
    vanishes, and $\Delta$ is negative within the region delimited by
    the red lines. The phase boundaries for lamellar--lamellar phase
    separations are marked by blue circles, solving
    eq.~(\ref{eq:commonT}), with interpolated blue lines. Black
    circles being \emph{within} the blue-enclosed region, unbinding is
    here circumvented by a lamellar--lamellar phase separation}
  \label{fig:phaseDiagramChiPhi}
\end{figure}

 Though a complete study is left to undaunted readers, a phase
diagram less schematic than in Fig.~\ref{fig:diaPhaChiPhi} is
nevertheless displayed for illustration purposes in
Fig.~\ref{fig:phaseDiagramChiPhi}.
It is interesting to observe that, for this specific case, the
unbinding transition no longer occurs: For \emph{large enough}
interactions between surfactant bilayers--as described by parameter
$\chi$--, hydrating the lamellar stack leads to a lamellar--lamellar
phase separation, the dilute lamellar phase being
\emph{unbound}. Below a critical value $\chi_c$ (close to 3.067 in
the present case), the lamellar phase can be continuously swollen by
water until infinite dilution. A decade-old controversy about swelling
of lecithin-based lamellar phases~\cite{LeNeveu1976,Helfrich1978},
qualitatively at least, might be settled by the above-described feature.
\section{Newton black films}
\label{NBF}
\subsection{Thermodynamic considerations}
\label{thermo2D}
Considering the structural analogy between soap films and lyotropic
lamellar phases, the ideas leading to the model free energy for
stacked bilayers, eq.~(\ref{eq:rafaelReduced}), may be used almost
directly for thermodynamic models of soap films and, more
specifically, of Newton black films. The composition variable $\phi$
now refers to the amount of ``solute'' in the aqueous channel, $\ell$
being related to its thickness, while $\Gamma$ remains the number of
surfactant molecules per unit \emph{monolayer} area. At contrast with
the case of stacked bilayers, i) the exchange of surfactant molecules
between the two monolayers and the aqueous channel is explicitly taken
into account now, and ii) the meaning of the ``swelling law'' has to
be reinterpreted. Assuming homogeneous and ideally flat, parallel
monolayers, that is to say choosing a ``film model'' according to the
terminology of Ref.~\cite{Radke2015}, a simple geometric description
of a volume $V\equiv\ell{\cal A}=N_wv_w+N_{\mathrm{tot}}v$ with $N$,
respectively $N_{\mathrm{tot}}-N$, surfactant molecules covering two
monolayers with area $\cal A$, resp. dispersed in the aqueous channel,
gives
\begin{equation}
  \label{eq:lever2D}
  \phi=\frac{\phi_0-2\frac{v}{\ell}\Gamma}{1-2\frac{v}{\ell}\Gamma}
\end{equation}
with $\phi_0=N_{\mathrm{tot}}v/(N_wv_w+N_{\mathrm{tot}}v)$ the overall
system composition. For an easier comparison with eq.~(\ref{eq:dil}),
an equivalent form of such a geometric ``dilution law'' is
\begin{equation}
  \label{eq:dilNBF}
  \ell\frac{\phi_0-\phi}{1-\phi}=2v\Gamma
\end{equation}
As it appears below, the ``$\ell$--$\Gamma$'' interplay in soap films
is much less intuitive than in lamellar phases, the dilution law being
more intricate.
\par An important difference between (macroscopic) soap films and
Newton black films should be mentioned here: For macroscopic soap
films, at the water--air interface of a Langmuir trough for instance,
there is no thinning and $\ell$ is \emph{fixed}, usually to a value
much larger than any molecular scales. At contrast, both $\ell$ and
$\phi$ (or $\Gamma$) are fluctuating quantities in Newton black films,
$\ell$ moreover being in the order of a \emph{molecular} scale. The
thickness $h$ of the aqueous channel may be defined such that $h{\cal
  A}=N_wv_w+\left(N_{\mathrm{tot}}-N\right)v$, and it is also a
fluctuating quantity, directly linked to monolayer coverage $\Gamma$:
\begin{equation}
  \label{eq:filmThickness}
   h=\ell-2v\Gamma
\end{equation}
The so-called ``thick limit''--without thinning--thus describes,
\emph{e.g.} Langmuir troughs, and thinning is a characteristic feature
of (common or Newton) \emph{black films}.
\par A legitimate purpose of any thermodynamic model of soap or Newton
black films is, given the amount of surface-active material $\phi_0$
and, if appropriate, the film aspect ratio $\ell$, to build equations
of state for surfactant remaining dispersed within the aqueous channel
or sub-phase, $\phi$, and film coverage, $\Gamma$, as well as to
describe surface tension $\gamma$, surface dilational modulus $E$,
etc. If thinning of the (black) film by border capillary suction is
taken into account, predicting the optimal aspect ratio
$\ell_{\mathrm{opt}}$ is also, of course, one of the quite important
targets of the model. Such approaches were attempted, and fulfilled,
many times along the years, with various viewpoints, see for
instance~\cite{Prins1967,Chatelier1996,Kralchevsky1999,Fainerman2002,Groot2010}.
\par A convenient starting point is the Gibbs free energy with the
following form
\begin{equation}
  \label{eq:gibbs2D}
  G=G_{\mathrm{channel}}(N_w,N_{\mathrm{tot}}-N)+{\cal
    A}g_{\mathrm{film}}(\Gamma,h)
\end{equation}
with $G_{\mathrm{channel}}$ the Gibbs free energy for the species
dispersed within the aqueous channel, $g_{\mathrm{film}}$ the Gibbs
free energy (surface) density for the two surfactant-covered
monolayers facing each other across the aqueous channel and, of
course, $N=2\Gamma{\cal A}$~\cite{ftnt1}. The $h$ dependence in the
surface contribution is meant to account for film--film interactions
(``disjoining pressure'') across the aqueous channel that cannot be
neglected for Newton black films.
\par Osmotic equilibrium for the surfactant that exchanges between
monolayers and the aqueous channel requires
\begin{equation}
  \label{eq:minimisation2D}
  -\mu+\frac{1}{2}\left[\frac{\partial g_{\mathrm{film}}}{\partial\Gamma}-2v\frac{\partial
      g_{\mathrm{film}}}{\partial h}\right]=0
\end{equation}
a minimisation equation somehow similar to
eq.~(\ref{eq:minimisation}), Section~\ref{rafael}, where $\mu$ is the
surfactant chemical potential in the aqueous channel for composition
$\phi$.
\par Applying the Gibbs-Duhem relation to the first term in the
right-hand side of eq.~(\ref{eq:gibbs2D}) yields
\begin{equation}
  \label{eq:gibbsDuhem2D}
  G=N_w\mu_w(\phi)+N_{\mathrm{tot}}\mu(\phi)+{\cal
    A}\left[g_{\mathrm{film}}(\Gamma,h)-2\mu(\phi)\Gamma\right]
\end{equation}
with therefore the surface tension of the film, defined by the
area-dependent term in the Gibbs free energy, given by:
\begin{equation}
  \label{eq:surfT}
  \gamma=g_{\mathrm{film}}(\Gamma,h)-2\mu(\phi)\Gamma
\end{equation}
\par If thinning is taken into account, $\ell$ is also a fluctuating
quantity as mentioned above and a mechanical equilibrium condition has
to be considered in addition to the osmotic equilibrium condition,
eq.~(\ref{eq:minimisation2D}). In the so-called ``thin film pressure
balance'' set-up (see Ref.~\cite{Mysels1966} and, more recently,
\emph{e.g.}, Ref.~\cite{Qu2007}), the hydrostatic pressure within the
film, $p_i$, differs from the pressure in the outside medium,
$p_e$. This amounts to adding a pressure term $\left[(p_e-p_i){\cal
    A}\right]\times\ell$ to the right-hand sides of
eq.~(\ref{eq:gibbs2D}) or (\ref{eq:gibbsDuhem2D}), see
Ref.~\cite{Radke2015}. Mechanical equilibrium of the film then
requires
\begin{equation}
  \label{eq:mechEq}
  \ell\left(\frac{\partial g_{\mathrm{film}}}{\partial
      h}+p_e-p_i\right)=g_{\mathrm{film}}-2\mu\Gamma
\end{equation}
\par In the ``thick film'' limit, the difference between $\phi$ and
$\phi_0$ disappears and the ``disjoining pressure'' contribution to
the film surface tension, though essential for ensuring an actual
\emph{thinning} of freely-suspended black films, becomes
meaningless. The classical relation
\begin{equation}
  \label{eq:gibbsIso}
  \frac{\partial\gamma}{\partial\ln\phi}=k_BT\times2\Gamma
\end{equation}
is then recovered (with a factor 2 because the film comprises two
facing monolayers) by differentiating eq.~(\ref{eq:surfT}), if it can
be assumed that the surfactant solution behaves ideally in the aqueous
sub-phase.
\par \emph{Global} stability of the (macroscopic) soap
film--\emph{i.e.} not yet considering local ``peristaltic''
fluctuations--can be studied by expanding the Gibbs free energy,
eq.~(\ref{eq:gibbs2D}), to second order with respect to fluctuations
$\delta\!\phi$ in aqueous channel composition or, equivalently,
fluctuations $\delta\Gamma$ in surface coverage, both originating in
$\delta\!N$ fluctuations. In terms of surface coverage fluctuations,
the result is
\begin{equation}
  \label{eq:surfE}
  G\approx
  G_{\mathrm{opt}}+\frac{\cal A}{2}\left[-\frac{\partial^2f_{\mathrm{exc}}}{\partial\phi^2}\frac{4v^2(1-\phi)^2}{\ell\left(1-\frac{2v}{\ell}\Gamma\right)}+
    \frac{\partial^2g_{\mathrm{film}}}{\partial\Gamma^2}-4v\frac{\partial^2g_{\mathrm{film}}}{\partial\Gamma\partial
      h}+4v^2\frac{\partial^2g_{\mathrm{film}}}{\partial h^2}\right]\delta\Gamma^2
\end{equation}
Global stability is thus ensured when the expression within square
brackets is positive. As a matter of fact, it is closely related to
the surface dilational modulus $E$, usually defined by
\begin{equation}
  \label{eq:dilatMod}
  E\equiv-\Gamma\frac{\delta\gamma}{\delta\Gamma}
\end{equation}
with however an ambiguity (also encountered when defining heat
capacities, for instance) regarding the mechanism for the
infinitesimal variation $\delta\Gamma$ in surface coverage. Here, the
dilational modulus is defined for \emph{fixed} geometric film
parameters $V\equiv N_wv_w+N_{\mathrm{tot}}v$ (incompressibility) and
$\cal A$ (no thinning), and is therefore given by
\begin{equation}
  \label{eq:dilatMod2}
  E=\left[\frac{\partial^2g_{\mathrm{film}}}{\partial\Gamma^2}-4v\frac{\partial^2g_{\mathrm{film}}}{\partial\Gamma\partial
      h}+4v^2\frac{\partial^2g_{\mathrm{film}}}{\partial
      h^2}\right]\Gamma^2
\end{equation}
that is to say \emph{without} the osmotic compressibility component
$\partial^2f_{\mathrm{exc}}/\partial\phi^2$ that appears in
eq.~(\ref{eq:surfE}).
\par
\begin{figure}[htb]
  \centering
  \includegraphics[width=0.95\columnwidth]{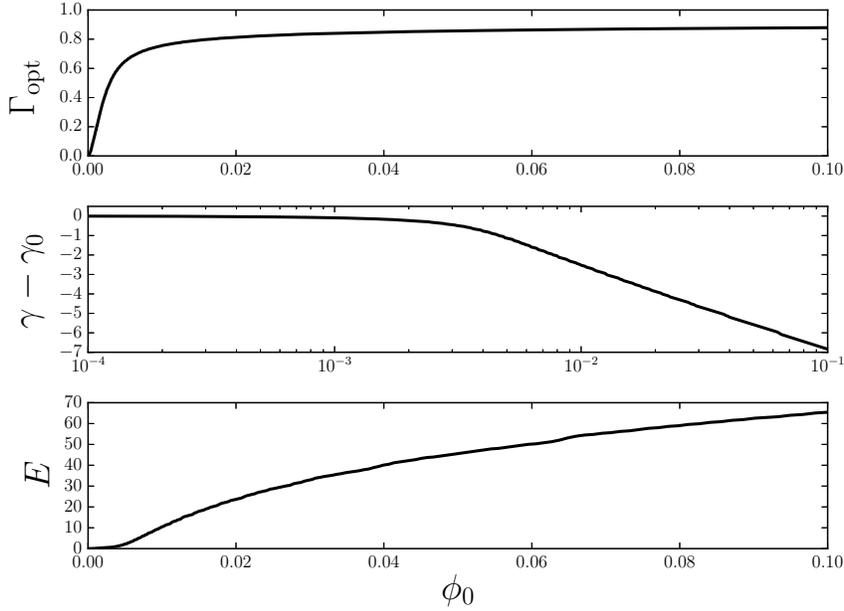}
  \caption{Equations of state in the ``thick film'' limit for optimal
    surface coverage $\Gamma_{\mathrm{opt}}$, surface tension $\gamma$
    and surface dilational modulus $E$, resulting from
    eq.~(\ref{eq:minimisation2D}), eq.~(\ref{eq:surfT}) and
    eq.~(\ref{eq:surfE}), respectively. Reduced units using
    $\Gamma_{\infty}$ and $k_BT\Gamma_{\infty}$ for surface coverage
    and surface energy, respectively}
  \label{fig:surfaceEOS}
\end{figure}

 As an illustration, optimal surface coverage,
$\Gamma_{\mathrm{opt}}$, surface tension $\gamma$ and surface
dilational modulus $E$ are plotted in Fig.~\ref{fig:surfaceEOS} as
functions of composition $\phi_0$ in the aqueous channel, with a
standard model for the chemical potential $\mu$--ideal solution:
$f_{\mathrm{exc}}=k_BT\left[(1-\phi)\ln(1-\phi)+\phi\ln\phi\right]/v$--and
a Gibbs free energy surface density $g_{\mathrm{film}}$ similar in
spirit to eq.~(\ref{eq:vdw2D}), but \emph{without} any ``disjoining
pressure'' contribution:
$g_{\mathrm{film}}=\gamma_0+k_BT\left\{\Gamma\ln\left[\Gamma/\left(\Gamma_{\infty}-\Gamma\right)\right]+b_2\Gamma^2/2-\chi\Gamma\right\}$,
where $\gamma_0$ is the water--air surface tension, $\Gamma_{\infty}$
the upper limit for surface coverage, $b_2$ a virial coefficient for
describing in-plane interactions between adsorbed surfactant molecules
and $\chi$ a measure of the adsorption energy per molecule.
Fig.~\ref{fig:surfaceEOS} gives a (model-dependent) quantitative and
explicit basis to the statement, adapted from
Ref.~\cite{Bergeron1996}:
\begin{quote}
  \textsf{It is also important to note, that since all of the
    film-stabilisation mechanisms rely on the same fundamental
    property, surfactant adsorption, they are inherently
    interdependent.}
\end{quote}
and, qualitatively at least, describes rather satisfactorily the
``thick film'' data~\cite{Bergeron1996}. Note that taking explicitly
into account the formation of surfactant micelles is not necessary to
explain a high value for the surface dilational modulus $E$.
\subsection{Peristaltic fluctuations}
\label{vrij}
The thinning phenomenon of a liquid film obviously requires $\ell$ to
vary. Moreover, as recalled in Section~\ref{intro}, the film stability
problem once an equilibrium value $\ell_{\mathrm{eq}}$ has been
reached should be investigated in the presence of (local) peristaltic
fluctuations. Owing to incompressibility, the overall volume
$V\equiv\ell{\cal A}$ cannot vary: Peristaltic fluctuations
($\delta\ell$ or $\delta\!h$) therefore imply area fluctuations
($\delta\!{\cal A}$), with surface tension acting as a ``restoring
force''. In addition, even though the total numbers of solvent $N_w$
and surfactant $N_{\mathrm{tot}}$ molecules remain constant, local
concentrations and surface coverage fluctuations are allowed and they
are, generally speaking, spatially dependent. Osmotic compressibility
in the aqueous channel and dilational modulus in the two monolayers
act as additional ``restoring forces''. The various fluctuating fields
are coupled, making a detailed analysis somehow involved.
\par Following \textsc{Vrij}, the stability analysis is here
restricted to the study of \emph{symmetric} modes with long wavelength
and small amplitude~\cite{Vrij1966}. It is also considered that the
thinning process has allowed to reach the value $\ell_{\mathrm{eq}}$
for the geometric parameter $\ell$. The film area
$V/\ell_{\mathrm{eq}}$ has therefore become a fixed quantity,
\emph{extremely large}, in molecular area units, for real black
films. With $t\equiv h/2$ half the \emph{local} thickness of the film,
a Gibbs free energy generalising eq.~(\ref{eq:gibbs2D}) and taking
into account conservation of matter may be expressed as
\begin{eqnarray}
  \label{eq:gibbs2D-local}
  G=\int
  d^2x&\left\{2t\left[p+(1-\phi)\frac{\mu_w^+}{v_w}+\phi\frac{\mu^+}{v}+f_{\mathrm{exc}}(\phi)\right]\right.
  \nonumber\\
  & \left.+\left[1+\frac{\left(\nabla
          t\right)^2}{2}\right]g_{\mathrm{film}}(\Gamma,t)\right\}
  \nonumber\\
  & -\frac{\mu}{v}\left(2t\phi-\ell\phi_0+2v\Gamma\right) \nonumber\\
  & -\frac{\mu_w}{v_w}\left[2t(1-\phi)-\ell(1-\phi_0)\right]
\end{eqnarray}
where $t(\mathbf{x})$, $\phi(\mathbf{x})$ and $\Gamma(\mathbf{x})$ are
now considered as fluctuating fields, $p$, $\mu_w^+$ and $\mu^+$ are
(constant) pressure and chemical potential reference parameters, and
$\mu$ (respectively, $\mu_w$) is a Lagrange multiplier ensuring the
conservation of $N_{\mathrm{tot}}$, resp. $N_w$, \emph{viz.}
conservation of the total number of surfactant, resp. solvent,
molecules. It is, of course, assumed here that the standard
thermodynamic description of the system can be safely extended down to
molecular scales.
\par The 3 Euler-Lagrange equations associated to the fluctuating
fields in eq.~(\ref{eq:gibbs2D-local}) are easily derived:
\begin{equation}
  \label{eq:eulerLagrange}
  \begin{array}{lcl}
    0&=&\left[1+\frac{(\nabla t)^2}{2}\right]\frac{\partial
      g_{\mathrm{film}}}{\partial\Gamma}-2\mu \\
    && \\
    0&=&-\frac{\nabla^2t}{2}g_{\mathrm{film}}+\left[1+\frac{(\nabla
        t)^2}{2}\right]\frac{\partial 
      g_{\mathrm{film}}}{\partial
      t} \\
    &&\mbox{}+2\left[p+(1-\phi)\frac{\mu_w^+-\mu_w}{v_w}+\phi\frac{\mu^+-\mu}{v}+f_{\mathrm{exc}}(\phi)\right]
    \\
    && \\
    0&=&-\frac{\mu_w^+-\mu_w}{v_w}+\frac{\mu^+-\mu}{v}+\frac{\partial
      f_{\mathrm{exc}}}{\partial\phi}
\end{array}
\end{equation}
with two conservation laws associated to the two Lagrange multipliers
\begin{equation}
  \label{eq:conservationLaws}
  \begin{array}{lcl}
    0&=&\int d^2x \left[2t\phi-\ell\phi_0+2v\Gamma\right]\\
    && \\
    0&=&\int d^2x \left[2t(1-\phi)-\ell(1-\phi_0)\right]
\end{array}
\end{equation}
As expected, \emph{constant} channel composition $\phi$ and thickness
$h$ verifying eq.~(\ref{eq:lever2D}) and eq.~(\ref{eq:filmThickness})
are possible solutions of eq.~(\ref{eq:eulerLagrange}) with a constant
surface coverage $\Gamma$, and eq.~(\ref{eq:minimisation2D}) is
recovered for such constant ``fields''.
\par The second order expansion of the Gibbs free energy,
eq.~(\ref{eq:gibbs2D-local}), with peristaltic fluctuations about the
constant solution of the Euler-Lagrange equations explicitly taken
into account, readily leads to
\begin{equation}
  \label{eq:peristalticFluct}
  \begin{array}{ll}
    \delta\!G\approx\frac{1}{2}\int d^2x&\left[2t\frac{\partial^2f_{\mathrm{exc}}}{\partial\phi^2}\delta\phi^2+(\nabla\delta
      t)^2g_{\mathrm{film}}\right.\\
    &\left.\mbox{}+\frac{\partial^2g_{\mathrm{film}}}{\partial\Gamma^2}\delta\Gamma^2+2\frac{\partial^2g_{\mathrm{film}}}{\partial\Gamma\partial
        t}\delta\Gamma\delta t+\frac{\partial^2g_{\mathrm{film}}}{\partial
        t^2}\delta t^2\right]    
  \end{array}
\end{equation}
a relation more conveniently expressed in Fourier space owing to
Parseval's identity, introducing mode expansions of the fluctuating
fields
\begin{equation}
  \label{eq:fourier}
  \psi(\mathbf{x})=\sum_{\mathbf{k}}\psi_{\mathbf{k}}e^{i\mathbf{k}\cdot\mathbf{x}}
\end{equation}
as
\begin{equation}
  \label{eq:peristalticFourier}
  \begin{array}{ll}
    \delta\!G=\frac{\cal A}{2}\sum_{\mathbf{k}}&\left[2t\frac{\partial^2f_{\mathrm{exc}}}{\partial\phi^2}\phi_{\mathbf{k}}\phi_{-\mathbf{k}}\right.\\
    &\mbox{}+(\frac{\partial^2g_{\mathrm{film}}}{\partial
      t^2}+k^2g_{\mathrm{film}})t_{\mathbf{k}}t_{-\mathbf{k}}\\
    &\left.\mbox{}+2\frac{\partial^2g_{\mathrm{film}}}{\partial\Gamma\partial
        t}\Gamma_{\mathbf{k}}t_{-\mathbf{k}}+\frac{\partial^2g_{\mathrm{film}}}{\partial\Gamma^2}\Gamma_{\mathbf{k}}\Gamma_{-\mathbf{k}}\right]
  \end{array}
\end{equation}
Quite similarly to Section~\ref{diaPha}, the stability of the film
with respect to diverging fluctuations requires characteristic
inequalities to be fulfilled, namely
\begin{equation}
  \label{eq:stability2D}
  \left\{
    \begin{array}{ccl}
      0&<&\frac{\partial^2g_{\mathrm{film}}}{\partial\Gamma^2} \\
      0&<&\frac{\partial^2f_{\mathrm{exc}}}{\partial\phi^2} \\
      0&<&\frac{\partial^2g_{\mathrm{film}}}{\partial\Gamma^2}\left(\frac{\partial^2g_{\mathrm{film}}}{\partial
          t^2}+k^2g_{\mathrm{film}}\right)-\left(\frac{\partial^2g_{\mathrm{film}}}{\partial\Gamma\partial
          t}\right)^2
\end{array}
    \right.
\end{equation}
Conversely, if
\begin{equation}
  \label{eq:filmDeterminant}
  \Delta\equiv\frac{\partial^2g_{\mathrm{film}}}{\partial\Gamma^2}\frac{\partial^2g_{\mathrm{film}}}{\partial
    t^2}-\left(\frac{\partial^2g_{\mathrm{film}}}{\partial\Gamma\partial
      t}\right)^2
\end{equation}
is negative, any mode with a wavelength larger than the critical
wavelength
\begin{equation}
  \label{eq:lambdaC}
  \lambda_c=2\pi\sqrt{\frac{g_{\mathrm{film}}\frac{\partial^2g_{\mathrm{film}}}{\partial\Gamma^2}}{-\Delta}}
\end{equation}
is bound to grow without limit. As was the case for the lamellar phase
stability problem, eq.~(\ref{eq:stability}) in Section~\ref{diaPha},
film stability thus appears to be driven by an interplay between
\emph{lateral} compressibility and ``dilution'' properties, the latter
again depending on interactions between surfactant-covered interfaces
across aqueous channels, eq.~(\ref{eq:mechEq}).
\par It is interesting to observe that the stability criterion
obtained here is \emph{not} exactly equivalent to \textsc{Vrij}'s
criterion, even in the special limit where the above-mentioned
interplay between lateral compressibility and ``dilution'' properties
does not exist because the cross-term
$\partial^2g_{\mathrm{film}}/\partial\Gamma\partial t$ is equal to
0. Indeed, the factor in front of $k^2$ in the last condition,
eq.~(\ref{eq:stability2D}), is \emph{not} the surface tension
$\gamma$, eq.~(\ref{eq:surfT}), but the Gibbs free energy (surface)
density $g_{\mathrm{film}}$, which results in a slightly different
equation for the critical wavelength, eq.~(\ref{eq:lambdaC}), as
compared to Ref.~\cite{Vrij1966}.
\par To proceed further, explicit models for
$g_{\mathrm{film}}(\Gamma,h)$ should be considered. As has been done
for lamellar stacks of bilayers in Section~\ref{rafael}, quite a
common approach amounts to considering separately the ``lateral''
equation of state--surfactant adsorption isotherms--and the
``vertical'' equation of state (``disjoining pressure'')
contributions, with in the latter case terms from van der Waals and
electrostatic forces usually discussed in details, as for instance by
\textsc{Sheludko} in a classical text,
Ref.~\cite{Sheludko1967}. However, because of
eq.~(\ref{eq:stability2D}), ``lateral'' and ``vertical'' contributions
should be considered simultaneously, at least in principle, as hinted
already long ago~\cite{Young1981}: On the basis of a numerical study
of van der Waals attractions indicating that the Hamaker constant
actually depends on surface coverage~\cite{Donners1977}, it was then
suggested that the optimal coverage $\Gamma_{\mathrm{opt}}$ not only
depends on $\phi_0$, but could also depend on film thickness even
though any ``significant dependence'' was ruled out~\cite{Young1981}.
\par Unfortunately, physically sound models do not seem to be
available yet for Newton black films. For nevertheless illustrating
the power of the above-described approach and, hopefully, motivating
further research, the study of a ``toy'' model with a reasonable
``lateral'' contribution and a (partially) \emph{ad hoc} ``vertical''
contribution is proposed below. Specifically
\begin{equation}
  \label{eq:toy}
  \begin{array}{ll}
    g_{\mathrm{film}}(\Gamma,h)=&\gamma_0\\
    &\\
    &\mbox{}+k_BT\left\{\Gamma\ln\left[\Gamma/\left(\Gamma_{\infty}-\Gamma\right)\right]+b_2\Gamma^2/2-\chi\Gamma\right\}\\
    &\\
    &\mbox{}+k_BT\Gamma_{\infty}h_{\mathrm{min}}/\left(h-h_{\mathrm{min}}\right)-V/h^2
  \end{array}
\end{equation}
with parameters $h_{\mathrm{min}}$ being the limiting thickness of a
wholly dehydrated Newton black film, and $V>0$ a Hamaker-like constant
associated to film--film van der Waals interactions across the water
channel. While the van der Waals term is, of course, extremely well
documented and not subject to controversy, it is important to stress
here that the repulsive term that \emph{diverges} at a finite
separation $h_{\mathrm{min}}$ in eq.~(\ref{eq:toy}) is here introduced
\emph{ad hoc}, not being supported by any physical argument similar to
\textsc{Helfrich}'s expression for the undulation interaction in
lamellar phases, eq.~(\ref{eq:helfrich}) in
Section~\ref{rafael}. Other choices would be equally acceptable if
they implement a very strong repulsion at small separations as, for
instance, an adaptation for the one-dimensional geometry relevant here
of the repulsive part of a $(6,12)$ \textsc{Lennard-Jones} potential,
yielding a $1/h^9$ term~\cite{Casteletto2003}. An electrostatic
repulsive contribution should actually be added to the ``vertical''
component in eq.~(\ref{eq:toy}), as it is practically always observed,
even with non-ionic surfactants, in disjoining pressure
measurements~\cite{Bergeron1996,Casteletto2003,Schlarmann2003}.
\par
\begin{figure}[htb]
  \centering
  \includegraphics[width=0.95\columnwidth]{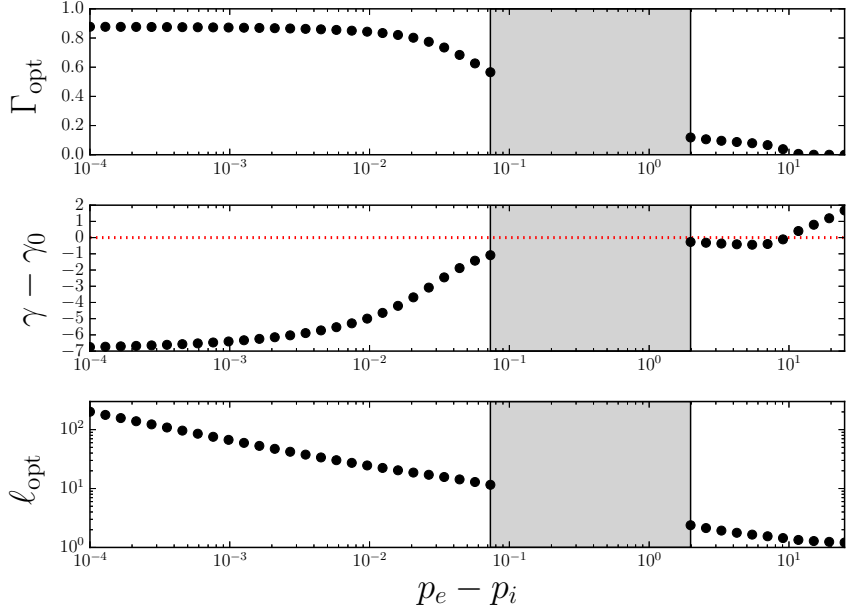}
  \caption{Black film equations of state in reduced units as functions
    of differential pressure $p_e-p_i$ for optimal surface coverage
    $\Gamma_{\mathrm{opt}}$, surface tension $\gamma$ and optimal
    thickness $\ell_{\mathrm{opt}}$, resulting from
    eq.~(\ref{eq:minimisation2D}), (\ref{eq:surfT}) and
    (\ref{eq:dilNBF}), respectively. The high (low) differential
    pressure range corresponds to a Newton (common) black
    film. Peristaltic fluctuations are divergent in the cray
    region. The horizontal dotted red line in the second panel
    indicates a surface tension equal to the bare water--air surface
    tension $\gamma_0$. Note that optimal surface coverage starts to
    decrease when approaching the unstable pressure range, and is
    \emph{much lower} for Newton black films than for common black
    films}
  \label{fig:vrij}
\end{figure}

 Fig.~\ref{fig:vrij} gives the results for black film coverage
$\Gamma$, surface tension $\gamma-\gamma_0$ and thickness $\ell$ , as
a function of differential pressure $p_e-p_i$ in the film obtained by
minimising the black film Gibbs free energy,
eq.~(\ref{eq:gibbsDuhem2D}) with the pressure term included. The grey
regions mark the pressure range where peristaltic fluctuations are
diverging, \emph{i.e.} $\Delta<0$ in eq.~(\ref{eq:filmDeterminant}).
\par
\begin{figure}[htb]
  \centering
  \includegraphics[width=0.95\columnwidth]{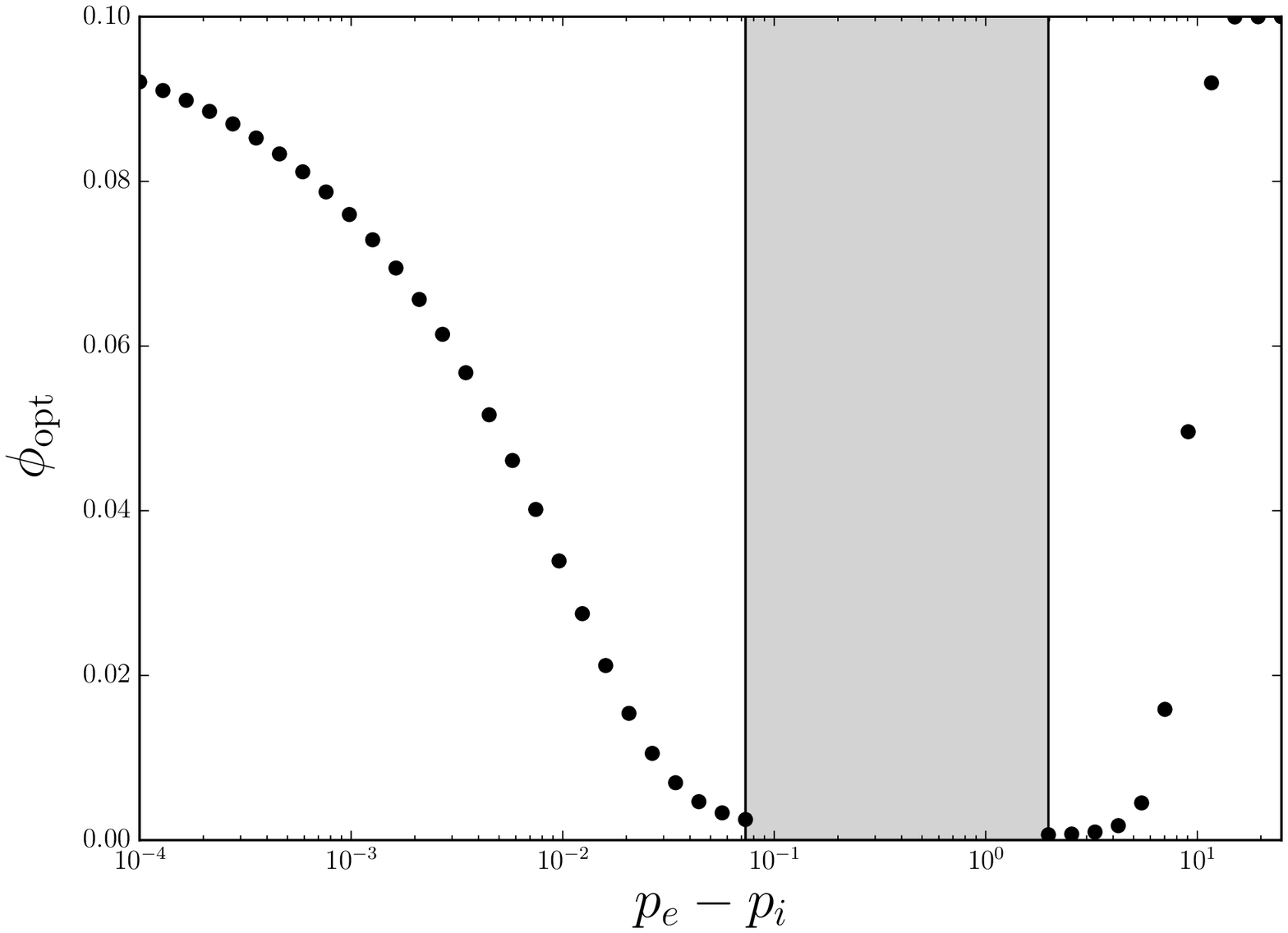}
  \caption{Optimal channel composition $\phi_{\mathrm{opt}}$ as a
    function of the differential pressure $p_e-p_i$}
  \label{fig:channelComposition}
\end{figure}

 It is interesting to note that the optimal surface coverage
predicted by the model varies quite significantly, and starts
\emph{decreasing} when enough differential pressure is applied to the
film or, in other words, when the film becomes
\emph{black}~\cite{ftnt2}. Such a feature, qualitatively speaking, is
not unexpected. Indeed, a 1~cm$^3$ volume of, say, sodium dodecyl
sulphate at the critical micellar concentration contains \emph{ca.}
$N_{\mathrm{tot}}=5\times10^{18}$ dodecyl sulphate, surface-active
ions that have to cover two air--water interfaces of area ${\cal
  A}=2\times10^2$~m$^2$ if the film aspect ratio is $\ell=5$~nm. The
upper limit for surface coverage is therefore
$\Gamma_{\mathrm{max}}=1.25\times10^{-2}$~nm$^{-2}$, nearly 3 orders
of magnitude less than the maximum surface coverage in the ``thick
film'' limit, typically a few molecules per squared
nanometres~\cite{Bergeron1996}. Owing to the osmotic equilibrium
between the two surfactant-covered interfaces and aqueous channel,
\emph{some} surface-active molecules have to remain dispersed in the
core part of the film, which leads to an optimal coverage even less
than $10^{-3}\times\Gamma_{\infty}$. Fig.~\ref{fig:channelComposition}
illustrates this intricacy of the ``$\ell$--$\Gamma$'' interplay, with
the channel composition $\phi$ as a third player specific to thinning
films, encapsulated by eq.~(\ref{eq:dilNBF}).
The saturation of $\phi_{\mathrm{opt}}$ to $\phi_0$ in the limit of a
very high differential pressures, that is to say for a highly
dehydrated Newton black film is, presumably, an artefact of the
\emph{ad hoc} component of the ``toy'' model considered here, as it
would amount to unrealistically small optimal surface coverages
$\Gamma_{\mathrm{opt}}$ owing to the ``$\ell$--$\Gamma$'' interplay.
\par
\begin{figure}[htb]
  \centering
  \includegraphics[width=0.95\columnwidth]{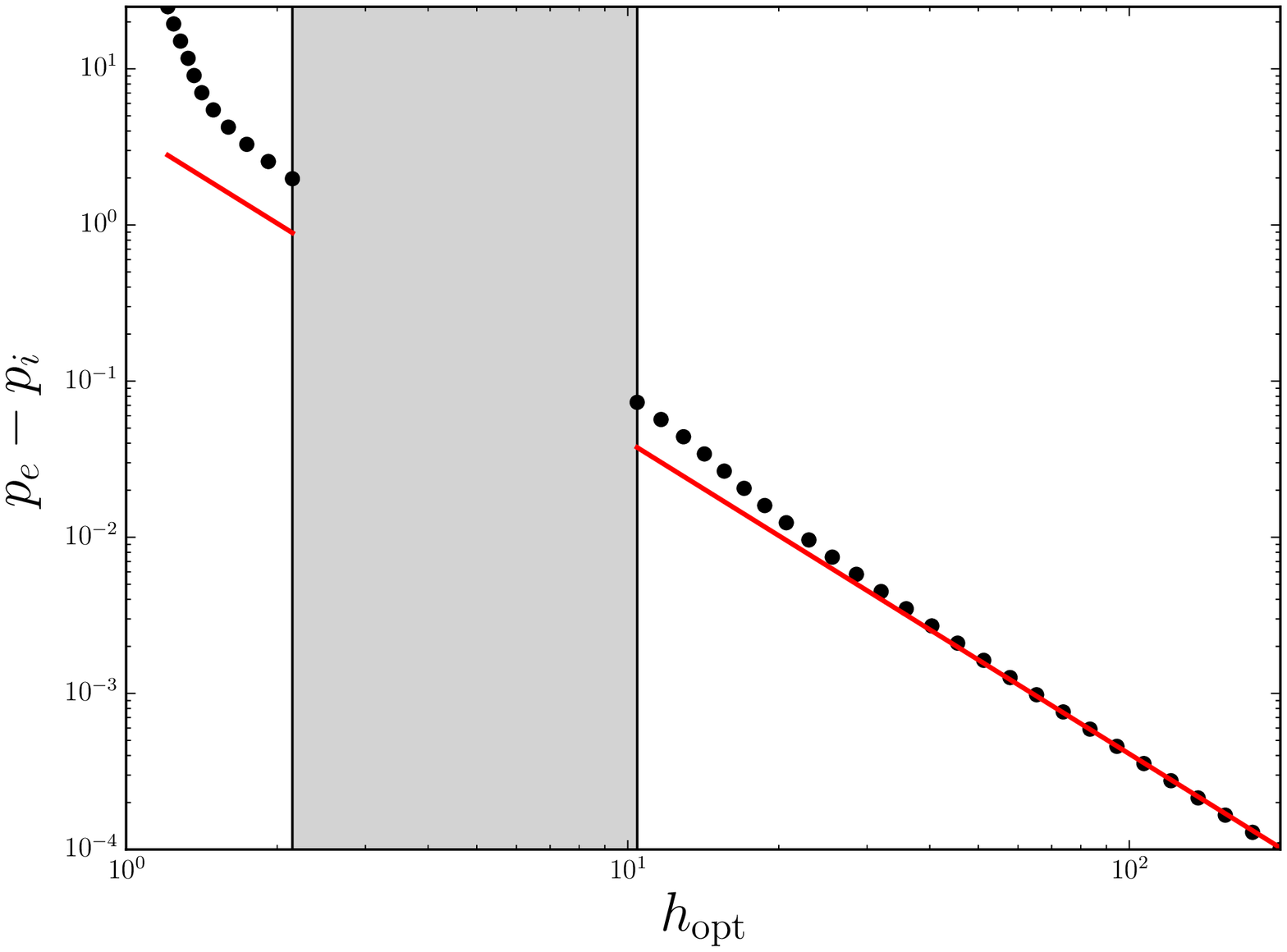}
  \caption{``Disjoining pressure''--black film core thickness equation
    of state in reduced units for the ``toy'' model,
    eq.~(\ref{eq:toy}). The red line corresponds to a power-law
    variation with exponent $-2$}
  \label{fig:myselsJones}
\end{figure}

 The equation of state for the black film ``disjoining pressure''
(identified to the differential pressure $p_e-p_i$ applied to it) as a
function of the black film aqueous core thickness $h$ is displayed in
Fig.~\ref{fig:myselsJones}.
Though using the common terminology is \emph{not} legitimate here,
because there is only \emph{one} repulsive barrier included in the
``toy'' model, eq.~(\ref{eq:toy}), it is tempting to describe as
Newton and, respectively common black films the domains located to the
left, resp. to the right, of the (grey) peristaltic instability
region. In the (kind of) common black film domain, the ``disjoining
pressure'' decays approximately as $h^{-2}$, \emph{viz.} varies as the
van der Waals contribution to the Gibbs free energy (surface) density
in eq.~(\ref{eq:toy}). The precise law for its divergence when
$h_{\mathrm{opt}}/h_{\mathrm{min}}\rightarrow1$ obviously depends on
the \emph{ad hoc} component of the ``toy'' model and should therefore
not be taken too seriously, even though it resembles, qualitatively at
least, to some data found
experimentally~\cite{Mondain-Monval1996,Schlarmann2003}.
\section{Conclusion}
\label{conclusion}
The structural analogy between soap films, in particular common or
Newton black films, and lamellar stacks of surfactant bilayers allows
considering these systems with a unified point of view regarding
fluctuations and interactions, on one hand, and thermodynamic
modelisation on the other hand. Phase diagrams, ``vertical'' or
``lateral'' equations of state have been routinely investigated for a
long time in the two kinds of systems, and a \emph{lingua franca} has
emerged from the shared problems and objectives in the film and
lamellar stack communities. The elusive nature of the short-range and
very strong repulsive interactions between surfactant-covered
interfaces, acting across water channels, remains controversial, but
both communities, warned long ago by visionary
precursors~\cite{Vrij1966,Helfrich1978}, are now firmly convinced that
fluctuations always play an essential role. A more in depth study of
the unavoidable coupling between ``lateral'' and ``vertical''
properties (or ``$\ell$--$\Gamma$ interplay'') may be a fruitful
direction to follow for a better understanding of the physics of thin
films and lamellar stacks. Some key assumptions in the field, namely
\begin{enumerate}
\item The various contributions to the disjoining pressure are
  additive (taken from~\cite{Bergeron1999}
\item Surface coverage deduced from pressure--area isotherms is
  constant over the range of Newton black film thickness (as
  formulated in Ref.~\cite{Jang2006})
\end{enumerate}
may have to be challenged for further progress.
\section*{Acknowledgements}
Though being, administratively speaking, a chemist, I was given the
opportunity to teach statistical physics in the physics department of
the Bordeaux university. Taking shamelessly inspiration from
Ref.~\cite{Radke2015a}, it is fair to say that
\begin{quote}
  \textsf{Some crucial elements of this teaching experience germinated
    (part of) this contribution. The challenge of teaching engenders
    learning.}
\end{quote}
I am grateful to prof. Philippe \textsc{Tamarat} and Brahim
\textsc{Lounis} for this challenging opportunity.
\bibliographystyle{unsrt}
\bibliography{../JabRef}{}

\begin{thebibliography}{10}
\bibitem{ftnt1}
It will be convenient here to consider the air--water interface of a {L}angmuir
  trough (``thick limit'') as splitted into two equivalent, half-area films.
\bibitem{ftnt2}
An aspect ratio parameter equal to 50~nm is close to $\ell=60$ in reduced units
  and is reached when the differential pressure exceeds
  $\approx1.3\times10^{-3}$--or about 10~k{P}a for the values chosen when
  drawing {F}ig.~\ref{fig:vrij}.
\end{thebibliography}
%
\end{document}